\documentclass{jfm}
\usepackage{graphicx}
\usepackage{epstopdf, epsfig}
\usepackage{amsmath,amssymb}
\shorttitle{Compressible HD turbulence}
\shortauthor{Hellinger et al}

\title{On cascade of kinetic energy in compressible hydrodynamic turbulence}

\author{Petr Hellinger\aff{1,2}
  \corresp{\email{petr.hellinger@asu.cas.cz}},
Andrea Verdini\aff{3,4},
  Simone Landi\aff{3,4}, 
Luca Franci\aff{5,4},
Emanuele Papini\aff{3,4}, and
Lorenzo Matteini\aff{6}
}

\affiliation{\aff{1}Astronomical Institute, CAS, Prague, Czech Republic
\aff{2}Institute of Atmospheric Physics, CAS, Prague, Czech Republic
\aff{3}Universit\`a di Firence, Italy
\aff{4}INAF, Osservatorio Astrofisico di Arcetri, Firenze, Italy
\aff{5}Queen Mary University of London, UK
\aff{6}Imperial College, London, UK
}

\begin{document}

\maketitle

\begin{abstract}
Properties of the turbulent cascade of kinetic energy are
studied using direct numerical simulations of
 three-dimensional hydrodynamic
decaying turbulence with a moderate Reynolds number and the
initial Mach number $M=1$.
Compressible and incompressible versions of
the K\'arm\'an-Howarth-Monin (KHM) and low-pass filtering/coarse-graining approaches
are compared. In the simulation the
total energy is well conserved; the scale dependent KHM
 and coarse-grained energy equations
are also well conserved; the two approaches show similar
results, the system does not have an inertial
range for the cascade of kinetic energy, the
region where this cascade dominates also have
a non-negligible contribution of the kinetic-energy decay, 
dissipation, and pressure-dilatation effects.
While the two approaches give semi-quantitatively similar results for the kinetic
energy cascade, dissipation and 
pressure-dilatation rates, they differ in the increment separation and filtering scales;
these scales are not simply related.
The two approaches may be used to find the inertial range and
to determine the cascade/dissipation rate of the kinetic energy.
\end{abstract}


\section{Introduction}

Turbulence in compressible fluids is not well understood. One
of the open questions is the existence of the so called inertial range, 
 where the kinetic energy cascades
(usually from large to small scales) without any losses. 
In the incompressible approximation hydrodynamic (HD) turbulence 
typically exhibits such inertial range provided that 
a large separation exists
between the driving/energy containing scales and the dissipation ones.
These properties are well described by the K\'arm\'an-Howarth-Monin (KHM)
equation \citep{kaho38,moya75} for statistically homogeneous turbulence.
This equation represents a scale-dependent energy conservation and relates the
driving/decay of kinetic energy, its cascade and dissipation. 
The inertial range can be formally defined as the region where the driving/decay and dissipation
are negligible and so that the dominant process is the cascade; this leads
in the infinite Reynolds number limit 
to so called exact (scaling) laws for isotropic media
\citep{kolm41b,fris95}.

In the case of compressible HD turbulence, the kinetic energy
and the internal energy are coupled via the dissipation as well as
through compressible (pressure dilatation) effects. In this
case it is not clear if there can be an inertial range of the
kinetic energy. One may consider the total (kinetic+internal) energy,
that is strictly conserved, but it is unclear if there is a
cascade of the total energy \cite[cf.,][]{eydr18}.
\cite{gaba11} derived the KHM equation for the total (kinetic and internal) energy 
assuming that the internal energy is governed by the isothermal closure.
This closure, however, partly decouples the internal and kinetic energies 
and does not conserve the total energy.
It is unclear if all or only a part of
pressure dilatation effects are present in such a system.
The cascade of the kinetic energy and 
pressure dilatation effects have 
not yet been studied in detail within the KHM approach.
On the other hand, the filtering/coarse graining approach
\citep{germ92,eyal09}
has been applied to the compressible turbulence \citep{alui11,alui13}
to derive relations equivalent to the KHM equation.
In particular, \cite{aluial12} show that the energy exchanges
between the kinetic and internal energies appear (at least for some
parameters) on large scales and
that there may exist a range of scales where the kinetic
energy cascades in a conservative way, forming an
 inertial range similar to that in the incompressible HD 
approximation. 
Here we reexamine the KHM equation
for the cascade of kinetic energy in compressible HD following \cite{gaba11},
we test it on results of numerical simulations
and compare these results with the coarse-graining approach.
The paper is organized as follows: in section~\ref{simulation} 
we present an overview of the direct 3D HD simulation 
with the initial Mach number $M=1$.
In section~\ref{cascade} we
 present the KHM equation for the kinetic energy for incompressible
and compressible HD and
we test these two versions
of KHM equation on the results of
the simulation.
In section~\ref{aluie}
we compare these results with the coarse-graining approach
assuming both incompressible and compressible approximations.
Finally, in section~\ref{discussion}
we discuss the results.

\section{Numerical simulation}
\label{simulation}
Here we use a 3D pseudo-spectral compressible hydrodynamic code derived from the compressible
MHD code \citep{verdal15} based on P3DFFT library \citep{peku12} and FFTW3
 \citep{frjo05}. 
The  code resolves 
the compressible Navier-Stokes equations, for the fluid density $\rho$,
 velocity $\boldsymbol{u}$, and the pressure $p$:
\begin{align}
\frac{\partial \rho}{\partial t}+ \boldsymbol{\nabla} \cdot (\rho \boldsymbol{u}) &= 0,
\label{density}\\
\frac{\partial (\rho\boldsymbol{u})}{\partial t}+ \boldsymbol{\nabla}\cdot (\rho \boldsymbol{u}\boldsymbol{u})
&=-\boldsymbol{\nabla}p
 +\boldsymbol{\nabla}\cdot\boldsymbol{\tau},
\label{velocity}
\end{align}
completed with an equation for the temperature $T=p/\rho$
\begin{align}
\frac{\partial T}{\partial t}+ (\boldsymbol{u} \cdot \boldsymbol{\nabla}) T =& \alpha \Delta T +
(\gamma-1) \left (-T \boldsymbol{\nabla}\cdot \boldsymbol{u} + \frac{1}{\rho}\boldsymbol{\nabla}\boldsymbol{u}:\boldsymbol{\tau}\right)
\label{temperature}
\end{align}
where  $\boldsymbol{\tau}$ is the viscous stress tensor
($\tau_{ij}=\mu\left(\partial u_{i}/\partial x_{j}+\partial u_{j}/\partial x_{i}-2/3\delta_{ij}\partial u_{k}/\partial x_{k}\right)$; here the dynamic viscosity $\mu$ is assumed to be constant)
and
 $\alpha$ is the thermal diffusivity (we set $\alpha=\mu$ and $\gamma=5/3$);
the colon operator denotes the double contraction of second order tensors,
$\boldsymbol{\mathrm{A}}:\boldsymbol{\mathrm{B}}=\sum_{ij}A_{ij} B_{ij}$. 
The box size is $(2\pi)^3$ (with a grid of $1024^3$ points), 
periodic boundary conditions are assumed.
The simulation is initialized with isotropic, random-phase, solenoidal 
fluctuations ($\boldsymbol{\nabla}\cdot\boldsymbol{u}=0$)
on large scales (with wave vectors $k=|\boldsymbol{k}|\leq 4$) having
the rms Mach number $M=1$ and a $k^{-1}$ 1-D power spectrum profile.
 We set the (constant) dynamic viscosity 
$\mu=2.8\ 10^{-3}$.

\begin{figure}
\centerline{\includegraphics[width=10cm]{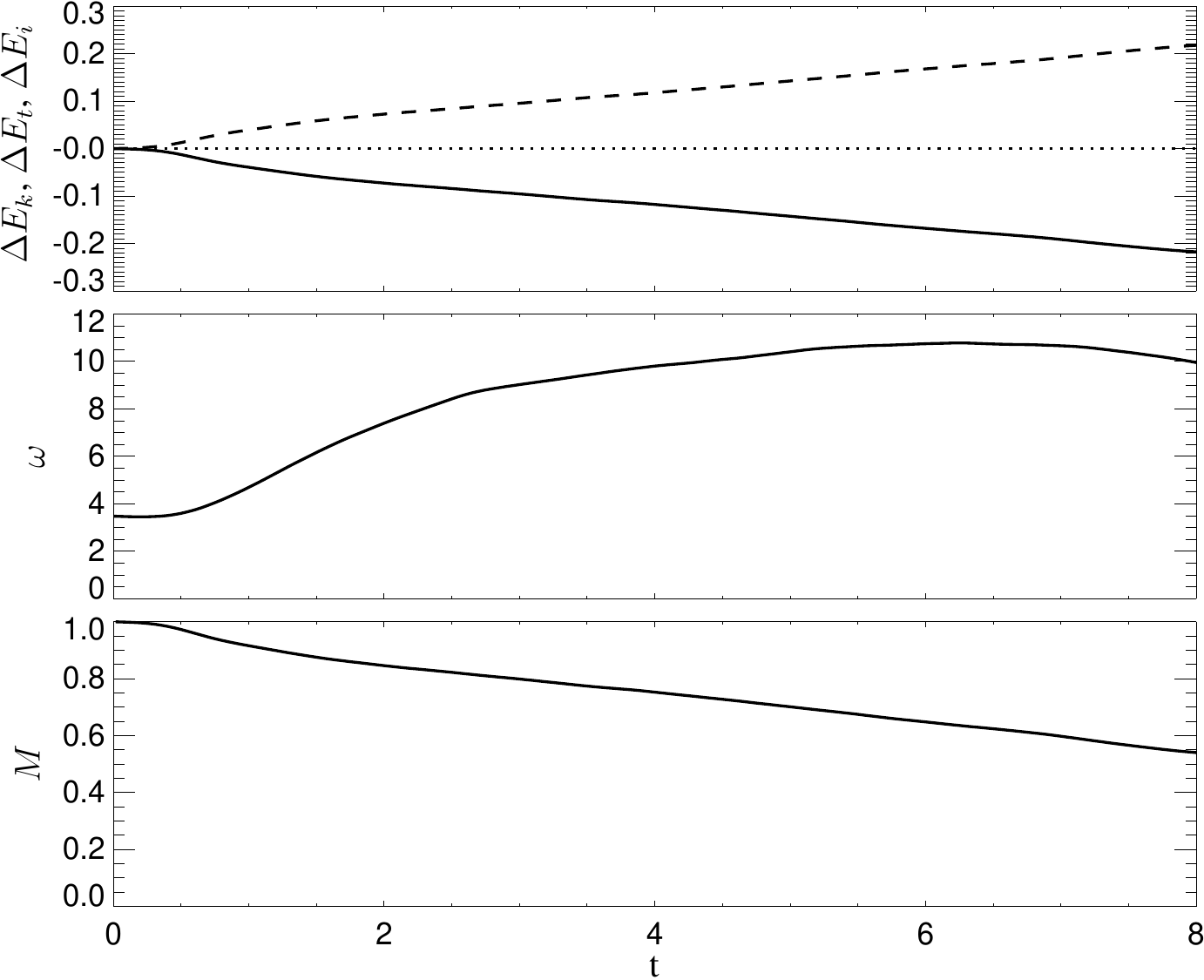}}
\caption{Evolution of (top) the relative changes 
in the kinetic energy $\Delta E_k$ (solid line), the total energy
 $\Delta E_t$ (dotted line), and the internal energy $\Delta E_i$ (dashed),
(middle) vorticity $\omega$, (bottom) Mach number $M$ as functions of time.
\label{evol}
}
\end{figure}

The evolution of the simulation is shown in Figure~\ref{evol}.
In the simulation the total energy $E_t=E_k+E_i$
 is well conserved. Here $E_k=\langle \rho u^2 \rangle/2$
is the kinetic energy and $E_i=\langle \rho T \rangle/(\gamma-1)$ is
the internal one (here $\langle \bullet \rangle$ denotes averaging over the simulation box). 
Top panel of Figure~\ref{evol}
displays the evolution of the relative changes in these energies,
$\Delta E_{k,i,t}=(E_{k,i,t}(t)-E_{k,i,t}(0))/E_t(0)$. The relative decrease
of the total energy is negligible, $\Delta E_t(t=8)\sim -4\ 10^{-6}$.
The middle panel of Figure~\ref{evol} shows the evolution
of the rms of the vorticity
$\boldsymbol{\omega}=\boldsymbol{\nabla}\times \boldsymbol{u}$.
The vorticity reaches a maximum at $t\simeq 6.2$; this is a signature
of a fully developed turbulent cascade.
The bottom panel of Figure~\ref{evol}
displays the evolution of the average Mach number $M$ (i.e., the
ratio between rms of the velocity and the mean sound speed). $M$ slowly decreases
during the evolution due to the decay of the level of fluctuations as
well as due to the turbulent heating that leads to
an increasing sound speed.

\begin{figure}
\centerline{\includegraphics[width=10cm]{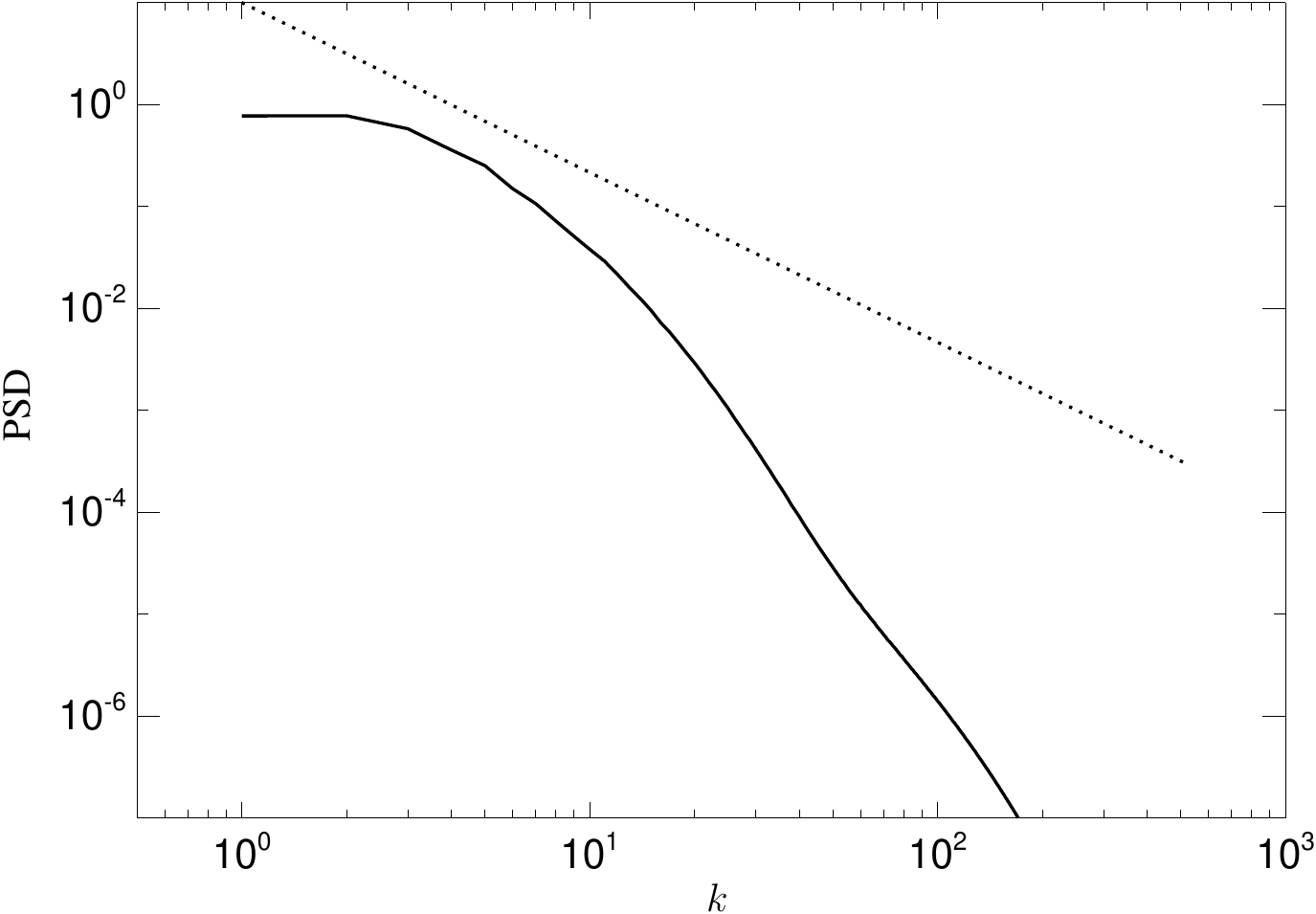}}
\caption{Power spectral density of $\boldsymbol{u}$ 
as a function of the wave vector $k$.
The dotted line denotes a dependence $\propto k^{-5/3}$.
\label{spec}
}
\end{figure}

Figure~\ref{spec} shows the power spectral density (PSD) of the velocity fluctuation
at the time $6.3$, around the maximum activity of
the vorticity, when turbulence is expected to be fully developed. The PSD does not exhibit
 a clear, Kolmogorov like
spectrum, thus suggesting that there is no inertial range in the simulation.
This is likely due to the small system size (small Reynolds number) and/or
due to the compressible effects. In the following sections we'll quantify these
effects using KHM and coarse graining approaches.

\section{KHM equation}
\label{cascade}
\subsection{Incompressible HD}
For the incompressible Navier-Stokes equation
\begin{align}
\frac{\partial\boldsymbol{u}}{\partial t}+ \boldsymbol{\nabla}\cdot (\boldsymbol{u}\boldsymbol{u})
&=-\frac{\boldsymbol{\nabla}p}{\rho}
 + \nu \Delta \boldsymbol{u},
\label{ivelocity}
\end{align}
where $\boldsymbol{u}$ is the velocity field, $\rho$ the density, $p$ the pressure,
$\nu$ is the kinematic viscosity. 
For statistically homogeneous decaying turbulence
one can get from Equation~(\ref{ivelocity}) the following form of the KHM
equation \citep{kaho38,moya75} in the terms of structure functions
of the increments of the velocity field $\delta\boldsymbol{u}=\boldsymbol{u}(\boldsymbol{x}+\boldsymbol{l})-\boldsymbol{u}(\boldsymbol{x})$
\begin{equation}
\frac{\partial S^{(i)}}{\partial t}+
\boldsymbol{\nabla}_{\boldsymbol{l}}\cdot \boldsymbol{Y}^{(i)}
 =  2 \nu \mathrm{\Delta}_{\boldsymbol{l}} S^{(i)}    - 4 \epsilon,
\label{KHM}
\end{equation}
where 
$S^{(i)}=\langle|\delta\boldsymbol{u}|^{2}\rangle$,
$\boldsymbol{Y}^{(i)}=\left\langle \delta\boldsymbol{u}|\delta\boldsymbol{u}|^{2}\right\rangle$
and
$\langle \bullet \rangle$ denotes statistical/spatial averaging 
($S^{(i)}$ and $\boldsymbol{Y}^{(i)}$ are functions of $\boldsymbol{l}$).
Equation~(\ref{KHM}) is simply related to the original form 
of the KHM equation that involves 
the cross-correlation $\left\langle\boldsymbol{u}(\boldsymbol{x}+\boldsymbol{l})
 \cdot\boldsymbol{u}(\boldsymbol{x})\right\rangle$
 \cite[cf.,][]{fris95}
\begin{align}
2\frac{\partial}{\partial t}\left\langle\boldsymbol{u}(\boldsymbol{x}+\boldsymbol{l})
 \cdot\boldsymbol{u}(\boldsymbol{x})\right\rangle 
-\boldsymbol{\nabla}_{\boldsymbol{l}}\cdot\boldsymbol{Y}^{(i)} =4\nu\Delta_{\boldsymbol{l}}\left\langle \boldsymbol{u}(\boldsymbol{x})\cdot\boldsymbol{u}(\boldsymbol{x}+\boldsymbol{l})\right\rangle 
\end{align}
since $S^{(i)}=2\langle |\boldsymbol{u}|^2\rangle -2\langle\boldsymbol{u}(\boldsymbol{x}+\boldsymbol{l})\cdot\boldsymbol{u}(\boldsymbol{x})\rangle$
and $\partial \langle |\boldsymbol{u}|^2\rangle/\partial t=-2\epsilon$.
Note that here the superscript $(i)$ denotes the incompressible approximation.
Equation~(\ref{KHM}) represents a scale-dependent energy-like conservation and relates the decay of kinetic energy
$\partial S^{(i)}/{\partial t}$,
the (incompressible)
dissipation rate (per mass) 
\begin{equation}
\epsilon= \nu \langle \boldsymbol{\nabla} \boldsymbol{u} : 
\boldsymbol{\nabla}\boldsymbol{u}  \rangle ,
\end{equation}
the cascade rate
$\boldsymbol{\nabla}_{\boldsymbol{l}} \cdot \boldsymbol{Y}^{(i)} $,
and the dissipation term $\nu \mathrm{\Delta}_{\boldsymbol{l}} \langle |\delta\boldsymbol{u}|^{2}\rangle$.
The inertial range can be formally defined as the region where the decay and dissipation terms
are negligible so that
\begin{equation}
\boldsymbol{\nabla}_{\boldsymbol{l}} \cdot \boldsymbol{Y}^{(i)} = - 4 \epsilon.
\label{inertial}
\end{equation}

For isotropic media, in the infinite Reynolds number limit, Equation (\ref{inertial}) 
leads to the  exact (scaling) laws \citep{kolm41b,fris95}.
Equation~(\ref{KHM}) is more general and may be directly tested 
in numerical simulations \cite[e.g.,][]{gotoal02} since
large Reynolds numbers needed for existence of the inertial range
are computationally challenging
 \cite[cf.,][]{ishial09}.

\subsection{Compressible HD}

Here we assume compressible Navier-Stokes equations,
Equations~(\ref{density},\ref{velocity}),
and investigate the structure function
$S=\left\langle \delta\boldsymbol{u}\cdot\delta\left(\rho\boldsymbol{u}\right)\right\rangle$
assuming a statistically homogeneous system following \cite{gaba11}.
After some manipulations (see appendix~\ref{apKHM} for details) we get

\begin{align}
\frac{\partial S}{\partial t}+\boldsymbol{\nabla}_{\boldsymbol{l}}\cdot\boldsymbol{Y}+R 
=C_p-C_{\tau}
	+2\left\langle \delta p\delta\theta\right\rangle - 2 \left\langle \delta\boldsymbol{\tau}:\delta\boldsymbol{\mathrm{\Sigma}}\right\rangle,
\label{KHMc}
\end{align}
where
$\boldsymbol{Y}=\left\langle \delta\boldsymbol{u}\left[\delta\left(\rho\boldsymbol{u}\right)\cdot\delta\boldsymbol{u}\right]\right\rangle$,
is a third-order structure function,
$\theta=\boldsymbol{\nabla}\cdot \boldsymbol{u}$ is the dilatation, 
$\boldsymbol{\Sigma}=\boldsymbol{\nabla}\boldsymbol{u}$ is the strain tensor, and 
$R=\left\langle \delta\boldsymbol{u} \cdot \left( \theta^\prime \rho \boldsymbol{u} -\theta \rho^{\prime}\boldsymbol{u}^{\prime}\right)\right\rangle $.

Here $C_p$ and $C_{\tau}$ are `correction' terms to $\left\langle \delta p\delta\theta\right\rangle$
and $\left\langle \delta\boldsymbol{\tau}:\delta\boldsymbol{\mathrm{\Sigma}}\right\rangle$,
respectively,
\begin{align}
C_p= \mathcal{C}\left[\boldsymbol{u},\boldsymbol{\nabla}p \right] 
\ \ \ C_{{\tau}}=\mathcal{C}\left[\boldsymbol{u},\boldsymbol{\nabla}\cdot\boldsymbol{\tau}\right], 
\label{correction}
\end{align}
where
\begin{align}
\mathcal{C}\left[\boldsymbol{a},\boldsymbol{b}\right]&=\left\langle \delta\boldsymbol{a}\cdot\delta\boldsymbol{b}-\delta\left(\rho\boldsymbol{a}\right)\cdot\delta\left(\frac{\boldsymbol{b}}{\rho}\right)\right\rangle 
=\left(\frac{\rho^{\prime}}{\rho}-1\right)\boldsymbol{a}^{\prime}\cdot\boldsymbol{b}+\left(\frac{\rho}{\rho^{\prime}}-1\right)\boldsymbol{a}\cdot\boldsymbol{b}^{\prime}.\nonumber
\end{align}
The $C_p$ and $C_{\tau}$ terms depend on the level of density fluctuations in the system.

The two terms, $S$ and $\boldsymbol{Y}$, are natural compressible generalization 
of $S^{(i)}$ and $\boldsymbol{Y}^{(i)}$, respectively. The $R$ term presents an additional
compressible energy-transfer channel \cite[cf.,][]{gaba11};
we do not see an obvious way how to turn this term to a divergence form
similar to $\boldsymbol{\nabla}_{\boldsymbol{l}}\cdot\boldsymbol{Y}$. The term
$\left\langle \delta p\delta\theta\right\rangle$ is a structure-function
formulation of the pressure dilation effect $p \theta$.
The viscous term 
$\left\langle \delta\boldsymbol{\tau}:\delta\boldsymbol{\Sigma}\right\rangle$
 corresponds 
to a combination of the two dissipation terms in the
incompressible case
$2\epsilon - \nu \Delta S^{(i)}$ in Equation~(\ref{KHM}).
On large scales, $|\delta \boldsymbol{x}|\rightarrow \infty$, where the correlations 
$\left\langle {\boldsymbol{\tau}(\boldsymbol{x}^\prime)}:\boldsymbol{\Sigma} \right\rangle\rightarrow 0$
the viscous term becomes twice the viscous heating rate
$Q_{\mu}$,
\begin{equation}
\left\langle \delta\boldsymbol{\tau}:\delta\boldsymbol{\Sigma}\right\rangle  \rightarrow 2 \left\langle \boldsymbol{\tau}:\boldsymbol{\Sigma}\right\rangle = 2 Q_{\mu}.
\end{equation}

Equation~(\ref{KHMc}) is analogous to Equation~(10) of \cite{gaba11} but it does not include
the isothermal internal energy assumed there (i.e., $p=c_s^2 \rho$, $e=c_s^2\ln{\rho/\rho_0}$, $c_s$: 
sound speed; see also appendix~\ref{apIE}). Also, in contrast with \cite{gaba11},
we do not consider forcing since we investigate decaying turbulence here.
Now we can test Equation (\ref{KHMc}) using the simulation results of section~\ref{simulation}.
We define the departure from zero of this equation as
\begin{align}
O(l) 
= \frac{1}{4} \left( -\frac{\partial S}{\partial t}- \boldsymbol{\nabla}_{\boldsymbol{l}}\cdot\boldsymbol{Y}
-  R  + 2\left\langle \delta p\delta\theta\right\rangle 
+C_p  -  2\left\langle \delta\boldsymbol{\tau}:\delta\boldsymbol{\mathrm{\Sigma}}
\right\rangle
-C_\tau \right).
\label{KHMO}
\end{align}
\begin{figure}
\centerline{\includegraphics[width=12cm]{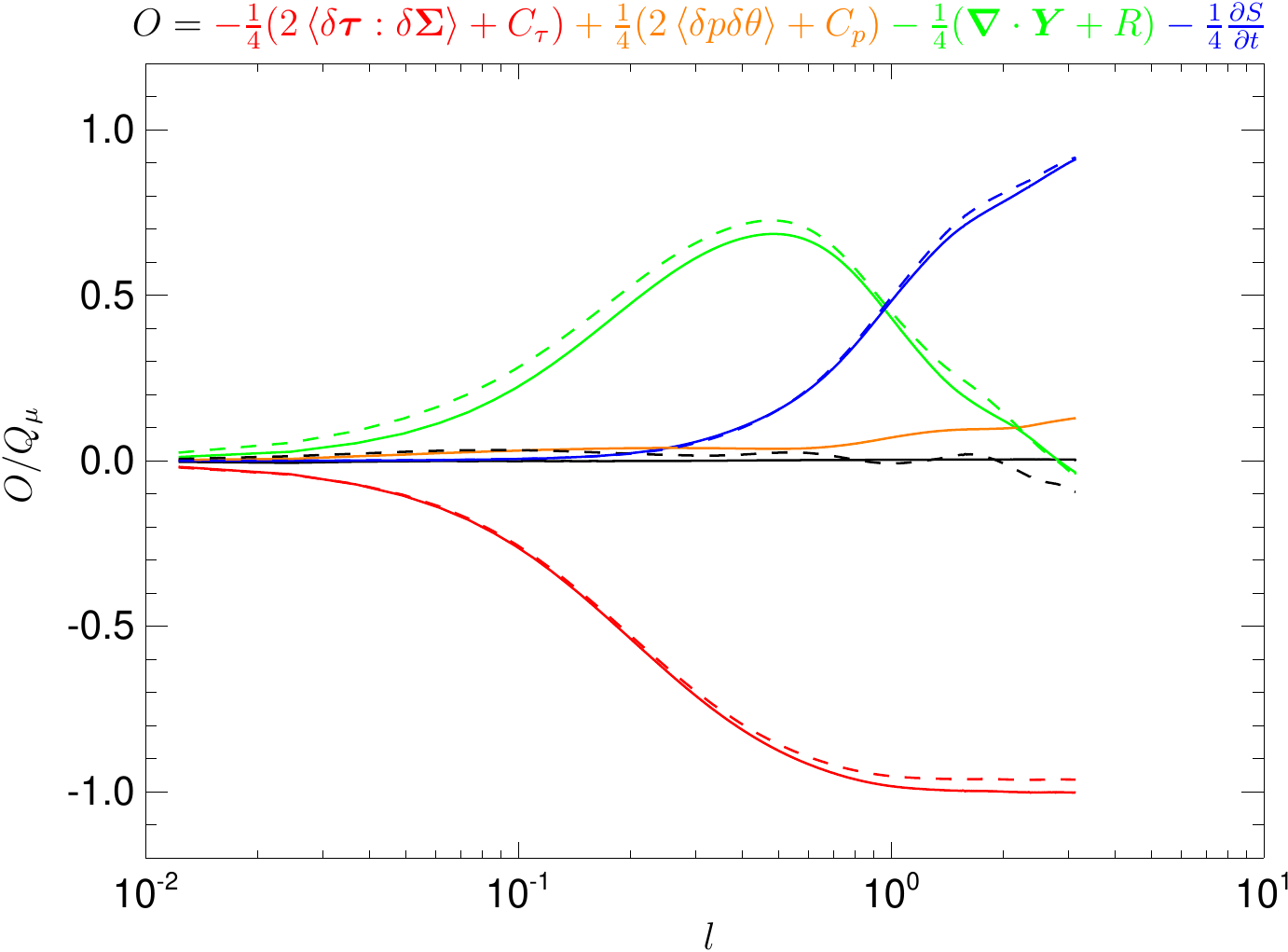}}
\caption{(black) The departure 
 $O$ (given by Equation~(\ref{KHMO})) as a function of the scale $l$ along with
the different contributions, the decaying term (blue) $-{\partial S}/{\partial t}/4$,
the cascade term (green) $-\boldsymbol{\nabla}_{\boldsymbol{l}}\cdot\boldsymbol{Y}/4-R/4$, the compressible coupling term (orange) $\left\langle \delta p\delta\theta\right\rangle/2+C_p/4$, and
(red) the scale-dependent dissipation term
$-\left\langle \delta\boldsymbol{\tau}:\delta\boldsymbol{\mathrm{\Sigma}}\right\rangle/2 -C_\tau$.
Dashed lines show the incompressible equivalent, (black) the departure $O^{(i)}$ 
(given by Equation~(\ref{KHMO})), the decaying term (blue) $-\rho_0{\partial S^{(i)}}/{\partial t}/4$,
the cascade term (green) $-\rho_0\boldsymbol{\nabla}_{\boldsymbol{l}}\cdot\boldsymbol{Y}^{(i)}/4$,
and  (red) the dissipation term
$\rho_0 \nu \mathrm{\Delta} S^{(i)}/2  - \rho_0\epsilon$.
$O$, $O^{(i)}$ and all their contributions are normalized to $Q_{\mu}$.
\label{yag}
}
\end{figure}
Figure~\ref{yag} shows (black) the departure
 $O$ as a function of the scale $l$ (isotropized/averaged over spherical angles) along with
the different contributions, the decaying term (blue) $-{\partial S}/{\partial t}/4$,
the cascade term (green) $-\boldsymbol{\nabla}_{\boldsymbol{l}}\cdot\boldsymbol{Y}/4-R/4$, 
the pressure dilation term
 (orange) $\left\langle \delta p\delta\theta\right\rangle/2+C_p/4$, 
and the scale-dependent dissipation term 
$ - \left\langle \delta\boldsymbol{\tau}:\delta\boldsymbol{\mathrm{\Sigma}}\right\rangle/2 -C_\tau $.
This calculation is done for times $6.2$ and $6.3$ (see Figure~\ref{evol}) over
a reduced box $512^3$ (taking every second point in all directions);
the structure functions are calculated over the full separation space and
isotropized/averaged over the spherical angles; the partial time
difference is approximated by the finite difference between the two times.
Figure~\ref{yag} demonstrates that the departure  $O$ is small
as predicted by Equation~(\ref{KHMc}); quantitatively we get $|O|/Q_\mu<0.006$. 

The decay, dissipation, and pressure dilatation
terms approach zero as $l\rightarrow 0$ and 
 reach their maximum absolute values on large scales:
the compressible dissipation term
$\left\langle \delta\boldsymbol{\tau}:\delta\boldsymbol{\mathrm{\Sigma}}\right\rangle/2
\rightarrow  Q_\mu $ as expected, and, 
similarly, $\partial S/\partial t /4 \sim
\partial \langle \rho |\boldsymbol{u}|^2 \rangle/\partial t /2\simeq 0.91 Q_\mu $
and $\langle \delta p \delta \theta \rangle/2 \sim \langle p \theta \rangle
\simeq 0.12 Q_\mu$. On large scales we recover the energy conservation
$\partial \langle \rho |\boldsymbol{u}|^2 \rangle/\partial t /2=-Q_\mu+
\langle p \theta \rangle$; the small error is likely due to the estimation
of the time derivative by the finite difference and other numerical effects.
The correction terms are small but not negligible
$|C_p|/Q_\mu/4<0.06$ and $|C_\tau|/Q_\mu/4<0.02$
and tend to zero on small and large scales.

 The cascade term is important on medium scales but
there is no true inertial range since the decay, dissipation as well as
the pressure dilatation term are not negligible there.
For larger Reynolds numbers one may expect that the
decay and pressure dilatation terms become negligible
on medium scales and that there is a range of scales
 the cascade term is compensated
by the constant dissipation term
\begin{align}
\boldsymbol{\nabla}_{\boldsymbol{l}}\cdot\boldsymbol{Y}+R=-4 Q_\mu,
\label{exactKHM}
\end{align}
i.e., the inertial range.

Figure~\ref{yag} also displays by dashed lines results of the corresponding
incompressible version of KHM equation, 
the departure from zero (renormalized by the background density $\rho_0$)
given by
\begin{equation}
 O^{(i)}(l)= \frac{\rho_0}{4} \left( -\frac{\partial S^{(i)}}{\partial t}-
\boldsymbol{\nabla}_{\boldsymbol{l}}\cdot\boldsymbol{Y}^{(i)}
 +  2 \nu \mathrm{\Delta} S^{(i)}  - 4\epsilon \right),
\label{KHMiO}
\end{equation}
The incompressible terms are comparable to their compressible
counterparts. In particular, the dissipation terms are close
to each other. This indicates that most of the dissipation is
incompressible. 
On the other hand, $\rho_0\boldsymbol{\nabla}_{\boldsymbol{l}}\cdot\boldsymbol{Y}^{(i)}$
and $\boldsymbol{\nabla}_{\boldsymbol{l}}\cdot\boldsymbol{Y}$  are almost identical,
the decrease of the cascade rate in the compressible KHM is
due to the compressible $R$ term.

\section{Coarse-graining approach}

\label{aluie}

Let us now compare the structure function approach with the coarse graining one.
This method is based on scale-dependent filtering of the compressible Navier-Stokes equation
\cite[cf.,][]{alui13}.
For any field $a(\boldsymbol{x})$ one defines a coarse-grained (low-pass filtered) 
field
\begin{equation}
\overline{a}_{\ell}(\boldsymbol{x})= \int_V G_{\ell}(\boldsymbol{r}) a(\boldsymbol{x}+\boldsymbol{r})\mathrm{d}^3\boldsymbol{r}
\end{equation}
where $G_{\ell}(\boldsymbol{r})$ is a convolution kernel, $\int_V G_{\ell}(\boldsymbol{r})\mathrm{d}^3\boldsymbol{r}=1$.
 Here we use a filter $G_{\ell}(\boldsymbol{r})=\ell^{-3} \mathcal{G}(\boldsymbol{r}/\ell)$
based on the kernel $\mathcal{G}(\boldsymbol{r})$ which has the following Fourier transform
\begin{equation}
\hat{\mathcal{G}}(\boldsymbol{k})\propto
\begin{cases}
\mathrm{exp}\left(- \frac{k^2}{1/4-k^2} \right) & k < 1/2\\
 0 & k\geq 1/2
\end{cases}
\end{equation}
where $k=|\boldsymbol{k}|$
  \cite[see][for details]{eyal09}.

To include the density variations one also defines, for each field $a(\boldsymbol{x})$, 
a density-weighted (Favre) filtered field
\begin{equation}
\tilde{a}_{\ell}(\boldsymbol{x})= \frac{ \overline{\rho a}_{\ell} (\boldsymbol{x}) }{\overline{\rho}_{\ell}(\boldsymbol{x})}.
\end{equation}

By applying the filtering to Equations~(\ref{density},\ref{velocity}) one gets
\begin{align}
\frac{\partial \overline{\rho}_{\ell}}{\partial t}+ \boldsymbol{\nabla} \cdot(\overline{\rho}_{\ell}\tilde{\boldsymbol{u}}_{\ell})&= 0,
\label{cgdensity}\\
\frac{\partial (\overline{\rho}_{\ell}\tilde{\boldsymbol{u}}_{\ell})}{\partial t}+ 
\boldsymbol{\nabla}\cdot (\overline{\rho}_{\ell} \tilde{\boldsymbol{u}}_{\ell}
\tilde{\boldsymbol{u}}_{\ell})
&= - \boldsymbol{\nabla}\cdot\left[ \overline{\rho}_{\ell} ( \widetilde{\boldsymbol{u}\boldsymbol{u}}_{\ell}-\tilde{\boldsymbol{u}}_{\ell}
\tilde{\boldsymbol{u}}_{\ell} )\right]   -\boldsymbol{\nabla}\overline{p}_{\ell}
 +\boldsymbol{\nabla}\cdot\overline{\boldsymbol{\tau}}_{\ell}.
\label{cgvelocity}
\end{align}

One can derive a filtered energy budget  to get
the following spatial averaged energy conservation equation (assuming a closed system) that removes the energy spatial transport
\begin{equation}
\frac{ \partial \langle \mathcal{E}_{\ell} \rangle}{\partial t} + \langle \Pi_{\ell} +\Lambda_{\ell} - \overline{p}_{\ell} \boldsymbol{\nabla}\cdot \overline{\boldsymbol{u}}_{\ell}  +  D_{\ell} \rangle=0 \label{aluiec}
\end{equation}
where $\langle \cdot \rangle$ denotes spatial averaging ($\langle a(\boldsymbol{x}) 
\rangle =\int_V a(\boldsymbol{x}) \mathrm{d}^3\boldsymbol{x}/V$) and

\begin{align}
\mathcal{E}_{\ell}&=\frac{1}{2} \overline{\rho}_{\ell}|\tilde{\boldsymbol{u}}_{\ell}|^2, \\
\Pi_{\ell} &= - \overline{\rho}_{\ell} \boldsymbol{\nabla}\tilde{\boldsymbol{u}}_{\ell}:(\widetilde{\boldsymbol{u}\boldsymbol{u}}_{\ell}-\tilde{\boldsymbol{u}}_{\ell}\tilde{\boldsymbol{u}}_{\ell}),\\
\Lambda_{\ell}  &= ( \tilde{\boldsymbol{u}}_{\ell} -\overline{\boldsymbol{u}}_{\ell} ) \cdot \boldsymbol{\nabla} \overline{p}_{\ell}, \\
D_{\ell} &= \boldsymbol{\nabla}\tilde{\boldsymbol{u}}_{\ell}: \overline{\boldsymbol{\tau}}_{\ell}.
\end{align}

Equation~(\ref{aluiec}) represents a coarse-graining equivalent to the KHM equation~(\ref{KHMc});
$\partial \mathcal{E}_{\ell}/{\partial t}$ describes the (scale-dependent)
kinetic energy decay, $\langle \Pi_{\ell}+\Lambda_{\ell}\rangle $ represents
the energy transfer across scales, $ \langle \overline{p}_{\ell} \boldsymbol{\nabla}\cdot \overline{\boldsymbol{u}}_{\ell} \rangle$
is the (scale-dependent) pressure dilatation term, and
$ \langle D_{\ell} \rangle$ is the dissipation term. 
Similarly one can get the incompressible version of Equation~(\ref{aluiec}) starting from
Equation (\ref{ivelocity}) \cite[cf.,][]{eyal09} as
\begin{equation}
\frac{ \partial \langle \mathcal{E}_{\ell}^{(i)} \rangle}{\partial t} + \langle \Pi_{\ell}^{(i)} + D_{\ell}^{(i)} \rangle=0, \label{aluiei}
\end{equation}
where 
\begin{align}
\mathcal{E}_{\ell}^{(i)}&=\frac{1}{2} \rho_0|\overline{\boldsymbol{u}}_{\ell}|^2, \\
\Pi_{\ell}^{(i)}&= - \rho_0\boldsymbol{\nabla}\overline{\boldsymbol{u}}_{\ell}:(\overline{\boldsymbol{u}\boldsymbol{u}}_{\ell}-\overline{\boldsymbol{u}}_{\ell}\overline{\boldsymbol{u}}_{\ell}),\\
D_{\ell}^{(i)} &= \mu \boldsymbol{\nabla}\overline{\boldsymbol{u}}_{\ell}:\boldsymbol{\nabla}\overline{\boldsymbol{u}}_{\ell}, 
\end{align}
and $\rho_0$ is the background density.

\begin{figure}
\centerline{\includegraphics[width=12cm]{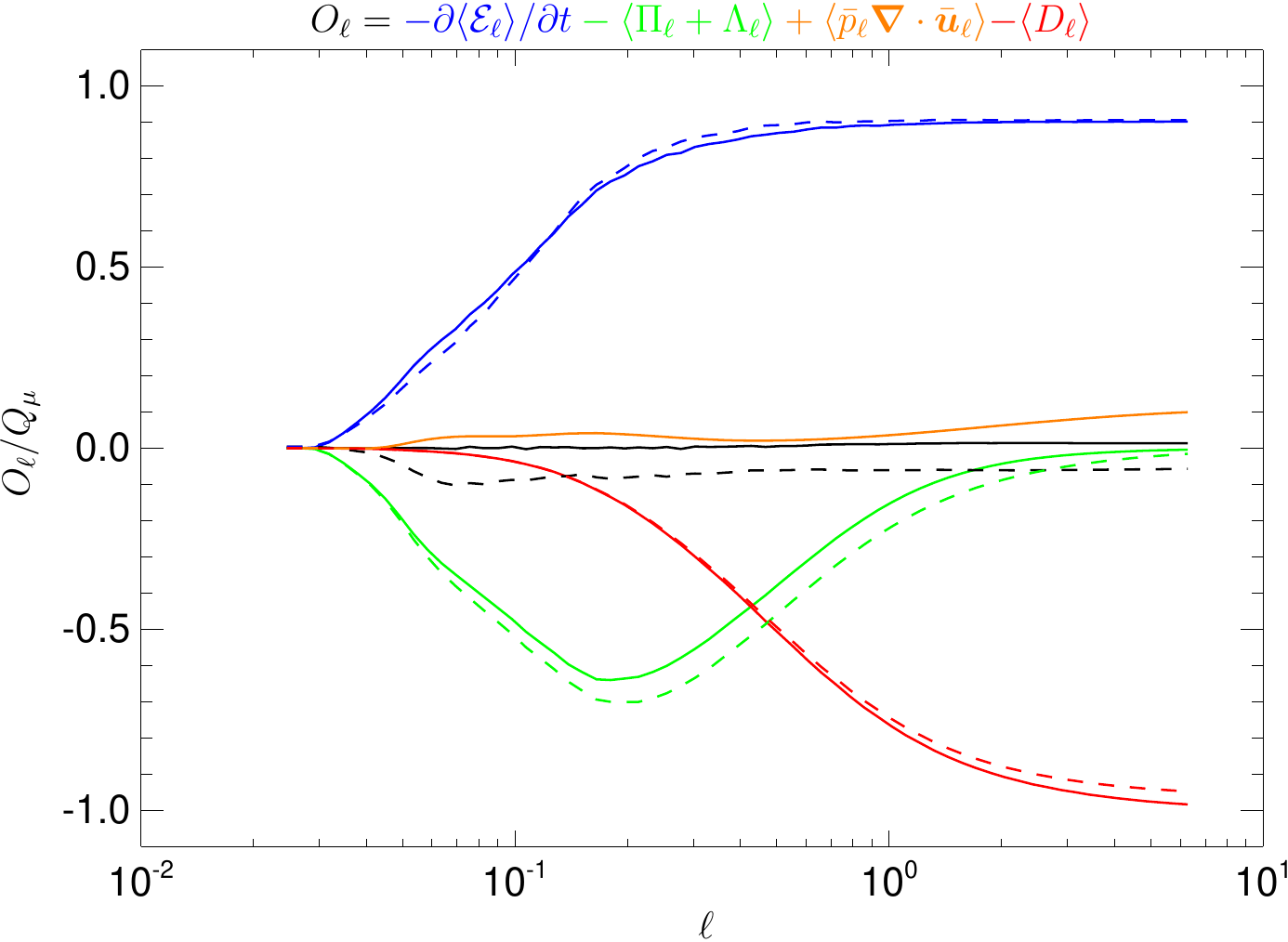}}
\caption{(solid) Departures from coarse-grained energy conservation
(black) $O_{\ell}$ (given by Equation~(\ref{testo})) as a function of the filtering 
scale $\ell$, along with
the different contributions: the decaying term (blue) $-\partial \mathcal{E}_{\ell}/{\partial t}$,
the energy transfer term (green) $\langle \Pi_{\ell}+\Lambda_{\ell}\rangle $, the 
pressure dilatation term (orange)
 $ \langle \overline{p}_{\ell} \boldsymbol{\nabla}\cdot \overline{\boldsymbol{u}}_{\ell} \rangle$,
and the dissipation term
$ \langle D_{\ell} \rangle$.
Dashed lines give the incompressible equivalents 
(black) $O_{\ell}^{(i)}$ (given by Equation~(\ref{testoi})), along with
the different contributions, the decaying term (blue) $-\partial \mathcal{E}_{\ell}^{(i)}/{\partial t}$,
the energy transfer term (green) $\langle \Pi_{\ell}^{(i)}\rangle$, 
and the dissipation term
$ \langle D_{\ell}^{(i)} \rangle$.
$O_{\ell}$, $O_{\ell}^{(i)}$ and all their contributions are normalized to $Q_{\mu}$.
\label{ccg}
}
\end{figure}

To test the validity  of Equation~(\ref{aluiec}), 
we define the departure from zero as
\begin{equation}
O_{\ell} = -\frac{ \partial \langle \mathcal{E}_{\ell} \rangle}{\partial t} - \langle \Pi_{\ell} +\Lambda_{\ell} - \overline{p}_{\ell} \boldsymbol{\nabla}\cdot \overline{\boldsymbol{u}}_{\ell}  +  D_{\ell} \rangle.
\label{testo}
\end{equation}
Figure~\ref{ccg} displays the results of the simulation of section~\ref{simulation} (solid lines), $O_{\ell}$ (normalized to $Q_{\mu}$) as a function of $\ell$ 
as well as the different contributions, the decaying term (blue) $-\partial \mathcal{E}_{\ell}/{\partial t}$,
the energy transfer term (green) $\langle \Pi_{\ell}+\Lambda_{\ell}\rangle $, the large scale pressure dilatation term (orange)
 $ \langle \overline{p}_{\ell} \boldsymbol{\nabla}\cdot \overline{\boldsymbol{u}}_{\ell} \rangle$,
and the dissipation term
$ \langle D_{\ell} \rangle$.
As in the KHM approach the calculation is done for times $6.2$ and $6.3$ over
a reduced box $512^3$.
Equation~(\ref{aluiec}) is in the simulation well satisfied, the departure $O_{\ell}$ is small, 
$|O_{\ell}|/Q_{\mu}\sim 10^{-2}$.

Figure~\ref{ccg} shows, similarly to the KHM results,
the decay, dissipation and pressure dilation terms go to
zero on small scales and on large scales they reach their
unfiltered counterparts:
$\langle D_{\ell}\rangle \rightarrow Q_\mu$, 
$\partial \mathcal{E}_{\ell}/\partial t \rightarrow
\partial \langle \rho|\boldsymbol{u}|^2 \rangle/\partial t /2$,
and 
$\langle\overline{p}_{\ell} \boldsymbol{\nabla}\cdot \overline{\boldsymbol{u}}_{\ell}\rangle
\rightarrow p \boldsymbol{\nabla}\cdot \boldsymbol{u} $.
The behaviors of the decay and dissipation terms are similar to their KHM
counterparts (see Figure~\ref{yag}) but the characteristic scales differ.
The pressure dilatation term is small but nonnegligible on all scales and overall decreases 
from large to small scales; this is also in agreement with the KHM results.
The energy transfer (cascade rate)
 $\langle \Pi_{\ell}+\Lambda_{\ell}\rangle $ is important on medium scales
and reaches a value comparable to that of the KHM cascade rate 
$-(\boldsymbol{\nabla}_{\boldsymbol{l}}\cdot\boldsymbol{Y}+R)/4$,
about $0.7 Q_{\mu}$; the main difference between the coarse graining 
and KHM results is the sign due to the different formulation of
the scale-dependent energy conservation.
A question is how the situation looks like for large Reynolds numbers
where there may be an inertial range. The present results suggest 
that in this case the cascade rate will be compensated by the
(constant) decay term in the inertial range
\begin{align}
\langle \Pi_{\ell} +\Lambda_{\ell}\rangle
 = - \frac{1}{2}\frac{ \partial \langle \rho|\boldsymbol{u}|^2 \rangle}{\partial t}.
\label{exactCG}
\end{align}

Figure~\ref{ccg} also shows (dashed lines) the results of the incompressible equivalent 
(see Equation~(\ref{aluiei})
\begin{equation}
O_{\ell}^{(i)} = -\frac{ \partial \langle \mathcal{E}_{\ell}^{(i)} \rangle}{\partial t} - \langle \Pi_{\ell}^{(i)} +  D_{\ell}^{(i)} \rangle.
\label{testoi}
\end{equation}
As in the KHM case, the incompressible decay and dissipation terms are close
to their compressible equivalents. 
The incompressible cascade rate $\langle \Pi_{\ell}^{(i)}\rangle$
is similar to that obtained for the incompressible KHM cascade. 
This support the interpretation
of $R$ as an additional compressible cascade in the
KHM equation.

\section{Discussion}
\label{discussion}
In this paper we investigated on the existence of the conservative cascade (inertial range)
of the kinetic energy in compressible hydrodynamic turbulence. We compared
the K\'arm\'an-Howarth-Monin and coarse-grained 
energy conservation approaches (in compressible and incompressible forms) 
for the kinetic energy, using data from a 3D HD
decaying turbulence simulation with a moderate Reynolds number and
the initial Mach number
 $M=1$.
In this simulation the two scale-dependent energy conservation equations
  are well satisfied. 
The pressure dilation coupling between the kinetic and internal energies
are the strongest on large spatial scales and decrease towards smaller
scales, in agreement with
the results of \cite{aluial12}.
Coherently with the PSD of the kinetic energy which does not show a Kolmogorov spectrum,
we do not observe a region where the kinetic-energy
cascade dominates, the effects of decaying, pressure dilation and dissipation being not negligible.
The KHM and coarse-graining approaches give 
rates of the cascade, decay, dissipation, and the pressure dilation
processes  that are in semi-quantitative agreement; the localization of
these different processes is, however, different when expressed 
in the scale separation or filtering spatial scales. This is not surprising,
calculations of structure functions and low-pass filtering
are very different procedures.  
The kinetic energy decay and dissipation rates estimated from
the incompressible approximations are close to the compressible
predictions.
In the simulation the observed kinetic-energy cascade is weaker than that predicted 
by the incompressible KHM equation, showing that the compressible term $R$ is not negligible.

In both approaches the pressure dilation terms $2\langle\delta p\delta\theta\rangle+C_p$ and $\langle \overline{p}_{\ell} \boldsymbol{\nabla}\cdot \overline{\boldsymbol{u}}_{\ell} \rangle$ seem to be weak in the region where the kinetic energy cascade term dominate:
We then expect that, depending on the level of compressibility, for a large enough Reynolds number \cite[cf.,][]{ishial09}
an inertial range for the kinetic energy may exist.
The pressure dilation effects typically weaken from large to 
small scales \citep{aluial12} and 
also inclusion of forcing extends the region where the cascade dominates.
The KHM and coarse-graining approaches could be used to
determine the heating/cascade rate.
The compressible equivalents of the incompressible ``exact'' 
laws, Equations~(\ref{exactKHM}) and \ref{exactCG},
have different meanings, the KHM approach gives the
(viscous) heating rate whereas the coarse graining
approach relates to the kinetic-energy decay rate
\cite[or to the energy injection rate in the 
forced turbulence, cf.,][]{alui13}.

In both the KHM and coarse-graining  approaches only the cascade of kinetic energy
is investigated. The effect
of including also the isothermal internal energy as used by \cite{gaba11} 
is questionable; the structure function $\langle \delta \rho \delta e \rangle$ proposed there
does not represent well the internal energy (see appendix~\ref{apIE}).
It is also questionable if a conservative cascade of
the kinetic energy exists
for strongly compressible (high Mach number) turbulence
\citep{eydr18,drey18}; an extension of this work to more compressible cases 
and/or larger Reynolds numbers is needed. 
The KHM structure function as well as coarse graining approaches
 may be further extended to (Hall) magnetohydrodynamics (MHD)
\citep{yangal17b,andral18,campal18,hellal18,ferral19} and even combined \citep{eyin03,kuzzal19}
to look at the localization of energy transfer processes.
One limitation of the usual coarse-graining approach is
that the filter is assumed to be isotropic; in anisotropic cases
(such as in rotating HD or magnetized MHD) an anisotropic filter may be more appropriate;
the KHM approach resolves this anisotropy rather naturally \citep{verdal15}.

\appendix

\section{Compressible K\'arm\'an-Howarth-Monin equation}
\label{apKHM}
Following \cite{gaba11}  we investigate the structure function 
$S=\left\langle \delta\boldsymbol{u}\cdot\delta\left(\rho\boldsymbol{u}\right)\right\rangle$.
To calculate $\partial S/\partial t$, we take  
Equation (\ref{velocity}) at two different points, $\boldsymbol{x}^\prime$ and $\boldsymbol{x}$,
and subtract them
\begin{equation}
\frac{\partial\delta\boldsymbol{u}}{\partial t}+(\boldsymbol{u}^{\prime}\cdot\boldsymbol{\nabla}^{\prime})\boldsymbol{u}^{\prime}	-(\boldsymbol{u}\cdot\boldsymbol{\nabla})\boldsymbol{u}=-\frac{\boldsymbol{\nabla}^{\prime}p^{\prime}}{\rho^{\prime}}+\frac{\boldsymbol{\nabla}p}{\rho}+\frac{1}{\rho^{\prime}}\boldsymbol{\nabla}^{\prime}\cdot\boldsymbol{\tau}^{\prime}-\frac{1}{\rho}\boldsymbol{\nabla}\cdot\boldsymbol{\tau}.
\label{a1}
\end{equation}
Here the primed variables are those at $\boldsymbol{x}^\prime$ (including $\boldsymbol{\nabla}^{\prime}=\boldsymbol{\nabla}_{\boldsymbol{x}^\prime}$). 
Similarly from a modified version of Equation (\ref{velocity}) we get
\begin{equation}
\frac{\partial\delta \left(\rho\boldsymbol{u}\right)}{\partial t}+
\boldsymbol{u}^{\prime}\cdot\boldsymbol{\nabla}^{\prime}\left(\rho^{\prime}\boldsymbol{u}^{\prime}\right)-
\boldsymbol{u}\cdot \boldsymbol{\nabla}\left(\rho\boldsymbol{u}\right)	=
-\left(\rho^{\prime}\boldsymbol{u}^{\prime}\right)\boldsymbol{\nabla}^{\prime}\cdot\boldsymbol{u}^{\prime}
+\left(\rho\boldsymbol{u}\right)\boldsymbol{\nabla}\cdot\boldsymbol{u}
-\boldsymbol{\nabla}^{\prime}p^{\prime} +\boldsymbol{\nabla}p
+\boldsymbol{\nabla}^{\prime}\cdot\boldsymbol{\tau}^{\prime}-\boldsymbol{\nabla}\cdot\boldsymbol{\tau}
\label{a2}
\end{equation}
Taking $\delta \left(\rho\boldsymbol{u}\right)$ times Equation~(\ref{a1}) plus $\delta \boldsymbol{u}$ times
Equation~(\ref{a2}) after some manipulation we have
\begin{align}
\frac{\partial S }{\partial t}
+ \boldsymbol{\nabla}_{\boldsymbol{l}}\cdot \left\langle \delta\boldsymbol{u} 
\left[\delta\left(\rho\boldsymbol{u}\right)\cdot\delta\boldsymbol{u}\right]\right\rangle 
=&-\left\langle (\boldsymbol{\nabla}^{\prime}+\boldsymbol{\nabla})\cdot\left[\boldsymbol{u}\delta\left(\rho\boldsymbol{u}\right)\cdot\delta\boldsymbol{u}\right]\right\rangle \nonumber \\
&+\left\langle \rho^{\prime}\boldsymbol{u}^{\prime}\cdot\delta\boldsymbol{u}\left(\boldsymbol{\nabla}\cdot\boldsymbol{u}\right)\right\rangle 
-\left\langle \delta\boldsymbol{u}\cdot\left(\rho\boldsymbol{u}\right)\left(\boldsymbol{\nabla}^{\prime}\cdot\boldsymbol{u}^{\prime}\right)\right\rangle 
\nonumber\\
&	-\left\langle \delta\boldsymbol{u}\cdot 
\delta \left( \boldsymbol{\nabla}p \right) \right\rangle
-\left\langle\delta\left(\rho\boldsymbol{u}\right)\cdot 
\delta \left( \frac{\boldsymbol{\nabla}p}{\rho}\right)\right\rangle
\nonumber \\
&	+\left\langle \delta\boldsymbol{u}\cdot
\delta\left(\boldsymbol{\nabla}\cdot\boldsymbol{\tau} \right) \right \rangle 
 +\left \langle \delta\left(\rho\boldsymbol{u}\right)\cdot
\left( \frac{\boldsymbol{\nabla}\cdot\boldsymbol{\tau}}{\rho} \right)\right\rangle 
\end{align}
The first term at the right hand side disappears in the homogeneous approximation 
and after some manipulation one gets Equation~(\ref{KHMc}).

\section{Internal Energy}
\label{apIE}

\cite{gaba11} investigated the KHM equation for the total energy
by representing the internal energy by the structure function
\begin{align}
S_e=\left\langle \delta\rho\delta e\right\rangle
\end{align}
where $e$ is the internal energy density. It is interesting to 
look at the properties of $S_e$ in the simulation of
section~\ref{simulation}. For $e$, 
 in our case $e=T/(\gamma-1)$,
one gets the following relation from Equation~(\ref{temperature})
\begin{align}
\frac{\partial e}{\partial t}+(\boldsymbol{u}\cdot\boldsymbol{\nabla})e=
\alpha\Delta e-\frac{1}{\rho}p\theta
+\frac{1}{\rho} \boldsymbol{\tau}: \boldsymbol{\Sigma}.
\end{align}
Using the same approach as in appendix~\ref{apKHM} one gets
the following KHM-like equation 
\begin{align}
\frac{\partial S_{e}}{\partial t}+\boldsymbol{\nabla}_{\boldsymbol{l}}\cdot\boldsymbol{Y}_{e}+R_{e}=\alpha\left\langle \delta\rho\delta\left(\Delta e\right)\right\rangle -\mathcal{D}\left(p\theta\right)+\mathcal{D}\left(\boldsymbol{\tau}:\boldsymbol{\Sigma}\right),
\label{eKHM}
\end{align}
where
\begin{align}
\boldsymbol{Y}_{e}&=\left\langle \delta\boldsymbol{u}\delta\rho\delta e\right\rangle, \\
R_{e}&=\left\langle \delta e\rho\boldsymbol{\nabla}^{\prime}\cdot\boldsymbol{u}^{\prime}-\rho^{\prime}\delta e\boldsymbol{\nabla}\cdot\boldsymbol{u}\right\rangle,
\end{align}
and 
\begin{align}
\mathcal{D}\left(a\right)&=\left\langle \left(1-\frac{\rho}{\rho^{\prime}}\right)a^{\prime}+\left(1-\frac{\rho^{\prime}}{\rho}\right)a\right\rangle 
\end{align}

In Equation~(\ref{eKHM}) $\boldsymbol{\nabla}_{\boldsymbol{l}}\cdot\boldsymbol{Y}_{e}+R_{e}$
represents the energy transfer connected with $S_e$. The r.h.s of Equation~(\ref{eKHM})
strongly depends on the density variation; for a constant $\rho$ this side is zero.
Combining Equation~(\ref{KHMc})
with Equation~(\ref{eKHM}) as  $\partial (S/2+S_e)/\partial t $
one recovers to a large extent the results of \cite{gaba11},
except for the isothermal closure (and the forcing term); 
pressure-dilation effects are in \cite{gaba11} transformed to
a contribution to the cascade term using the
isothermal closure

To test Equation~(\ref{eKHM}) on the simulation results of section~\ref{simulation}, 
we define the departure (see section~\ref{cascade}) as
\begin{align}
O_e= \frac{1}{2}\left[-\frac{\partial S_{e}}{\partial t}-
\boldsymbol{\nabla}_{\boldsymbol{l}}\cdot\boldsymbol{Y}_{e}-R_{e}
+\alpha\left\langle \delta\rho\delta\left(\Delta e\right)\right\rangle 
-\mathcal{D}\left(p\theta\right)+\mathcal{D}\left(\boldsymbol{\tau}:\boldsymbol{\Sigma}\right) \right].
\label{oe}
\end{align}
\begin{figure}
\centerline{\includegraphics[width=12cm]{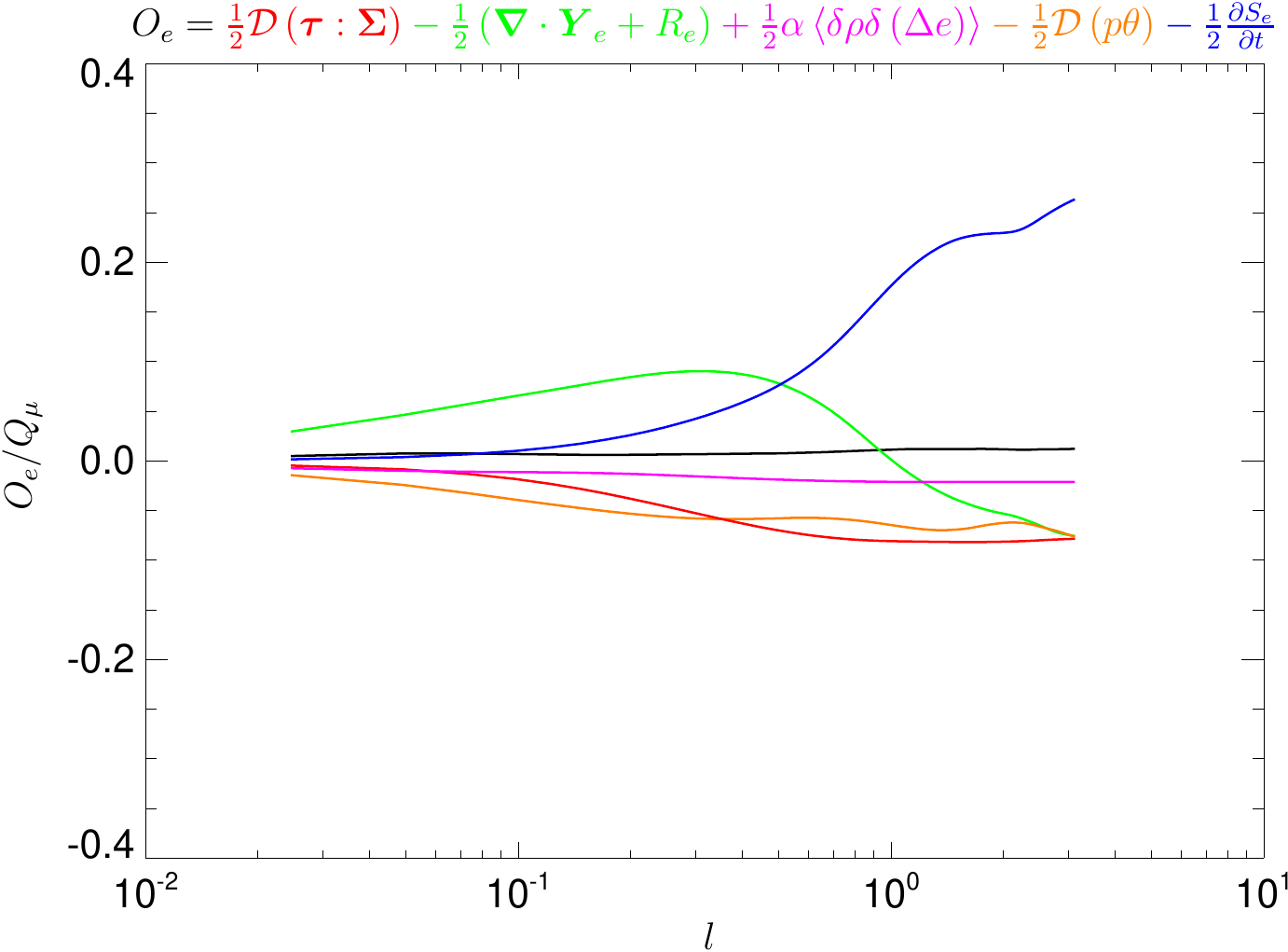}}
\caption{(black) The departure
 $O_e$ (given by Equation~(\ref{oe})) as a function of the scale $l$ along with
the different contributions, the decaying term (blue) $-{\partial S_e}/{\partial t}/2$,
the energy transfer term (green) $-(\boldsymbol{\nabla}_{\boldsymbol{l}}\cdot\boldsymbol{Y}_e+R_e)/2$, 
the pressure dilatation term (orange) $-\mathcal{D}\left(p\theta\right)/2$, and
(red) the dissipation term
$\mathcal{D}\left(\boldsymbol{\tau}:\boldsymbol{\Sigma}\right)$,
and the diffusion term (magenta) $\alpha\left\langle \delta\rho\delta\left(\Delta e\right)\right\rangle$.
$O_e$ and all its contributions are normalized to $Q_{\mu}$.
\label{eyag}
}
\end{figure}
Figure~\ref{eyag} shows the (isotropized) departure (black) $O_e$ as a function of the scale $l$ along with
the different contributions, the decaying term (blue) $-{\partial S_e}/{\partial t}/2$,
the energy transfer term (green) $-(\boldsymbol{\nabla}_{\boldsymbol{l}}\cdot\boldsymbol{Y}_e+R_e)/2$,
the pressure dilatation term (orange) $-\mathcal{D}\left(p\theta\right)/2$, and
(red) the dissipation term
$\mathcal{D}\left(\boldsymbol{\tau}:\boldsymbol{\Sigma}\right)$,
and the diffusion term (magenta) $\alpha\left\langle \delta\rho\delta\left(\Delta e\right)\right\rangle$.
The calculation is done on a sub-grid of $256^3$ points taking every fourth point in all directions
Figure~\ref{eyag} confirms that the energy-like conservation, Equation~(\ref{eKHM}),
is well satisfied, $|O_e|/Q_\mu\sim1$~\% and shows that all the terms are nonnegligible,
including the diffusion.

One important thing to note from Figure~\ref{eyag} is that ${\partial S_e}/{\partial t}<0$.
The structure function $S_e=\left\langle \delta\rho\delta e\right\rangle$ decreases
with the time in contrast with the internal energy $E_i=\langle \rho e \rangle$
that increases (see Figure~\ref{evol}).
If $S$ and $S_e$ are to represent kinetic and internal energy, respectively,
in an analogous way, the latter should increase as the former
decreases.
 This is a clear indication that $S_e$ does not
well represent the internal energy. Consequently,
the terms $\boldsymbol{\nabla}_{\boldsymbol{l}}\cdot\boldsymbol{Y}_e+R_e$ are not clearly related to
a cascade/energy transfer of the internal energy.

\acknowledgments
P.H. acknowledges grant 18-08861S of the Czech Science Foundation.

Declaration of Interests. The authors report no conflict of interest.’

\end{document}